\documentclass{aastex}
\usepackage{emulateapj5}
\usepackage{apjfonts}

\def\yr{{\rm yr}}

\def\kms{{\rm km}\,{\rm s}^{-1}}

\begin{document}

\submitted{To appear in ApJL}
\title{LSR0602+3910 -- Discovery of a Bright Nearby L-type Brown
Dwarf}

\author{Samir Salim\altaffilmark{1,3}, S\'ebastien
 L\'epine\altaffilmark{2}, R. Michael Rich\altaffilmark{1,3} \and
 Michael M. Shara\altaffilmark{2}}

\altaffiltext{1}{Department of Physics and Astronomy, University of
California at Los Angeles, Los Angeles, CA 90095, USA;
samir@astro.ucla.edu, rmr@astro.ucla.edu}

\altaffiltext{2}{Department of Astrophysics, Division of Physical
Sciences, American Museum of Natural History, Central Park West at
79th Street, New York, NY 10024, USA; lepine@amnh.org, shara@amnh.org}

\altaffiltext{3}{Visiting Astronomer, Lick Observatory}

\begin{abstract}

We report the discovery of LSR0602+3910, an L dwarf of class L1. The
object was initially identified by L\'epine et al.\ (2002) as a new
high proper motion star lying close to the Galactic plane. Its 2MASS
$J-K_s=1.43$ is consistent with an L dwarf, which we now confirm
spectroscopically. In addition, we see a signature of Li I absorption,
making LSR0602+3910 a brown dwarf, one of the brightest known
($K_s=10.86$). Among L dwarfs it is second in brightness to the
combined light of 2MASS 0746+20, a close binary system. We see no
indication that LSR0602+3910 is a binary, although high-resolution
imaging will be required to confirm this. Spectroscopic and
photometric distance estimates agree very well, placing LSR0602+3910
at $d=10.6\pm0.8$ pc.  LSR0602+3910 was most likely missed in previous
searches because of its proximity to the plane, the region that most
searches avoided. We estimate that some 40\% of bright L dwarfs are
missed because of this selection effect.

\end{abstract}


\section{Introduction}

The final decade of the last century has seen the discovery of
ultra-cool dwarfs. The low temperatures of these objects cause their
spectra to appear distinct from those of the latest M dwarfs. Thus the
new spectral class ``L'' was introduced \citep{k99,martin} in which
TiO and VO give way to lines of neutral alkalis and hydrides. Yet
cooler objects exhibit methane absorption, and are classed as ``T'' or
methane dwarfs.

The main question was to establish if any of these cool dwarfs was
substellar, i.e. whether they represented the long sought-after brown
dwarfs. The critical parameter -- mass -- is not directly observable,
so it must be derived from the stellar evolution models
\citep{burrows,baraffe} as a function of object's temperature and
age. Temperature is derived from spectra, while the age can sometimes
be independently estimated if the cool dwarf is a member of a binary
or a cluster. While all T dwarfs are considered too cool to sustain
hydrogen burning, L dwarfs can be either stars or brown dwarfs. Most
lines of evidence point to the conclusion that L dwarfs later than L4
are brown dwarfs (e.g. \citealt{gizis00}). For the earlier L dwarfs,
as for several very late M dwarfs, the presence of lithium can serve
as an indicator of substellar nature \citep{rebolo}, since it is
destroyed in objects having masses $>0.06 M_\sun$.

From the observational point of view, the recent discovery of a large
number of L dwarfs (250 are listed in D.\ Kirkpatrick's
database\footnote{\url{http://spider.ipac.caltech.edu/staff/davy/ARCHIVE/}}
updated on 2002 Dec 19) and T dwarfs (40 are listed in Burgasser's
archive\footnote{\url{http://www.astro.ucla.edu/~adam/homepage/research/tdwarf/index.html}}
updated on 2002 Dec 22), was facilitated by large area sky surveys --
primarily 2MASS and DENIS in the infrared, and SDSS in the optical. L
dwarfs are characterized by red infrared colors and faint or invisible
optical counterparts, while the T dwarfs have quite blue infrared
colors, but are also very faint or invisible in optical surveys. This
reversal of infrared colors between L and T dwarfs makes infrared
surveys inappropriate for the search for the L--T ``transitional''
objects. The reversal does not affect red optical bands like $i'$ and
$z'$, enabling SDSS to fill this gap \citep{leggett}. ``Transitional''
objects are now considered to be early T dwarfs \citep{burg,geballe}.
Yet other L dwarfs are found in targeted searches for companions to
nearby stars \citep{gizis01,k01,wilson}. Finally, L dwarfs are found
in new proper motion surveys \citep{lodieu,scholz}.

The L dwarf we present in this paper, LSR0602+3910, was first detected
as a new high proper motion object. Its discovery is described in
\S\ref{sec:disc}, and the observations confirming its nature are
described in \S\ref{sec:obs}. In \S\ref{sec:ana} we analyze and
interpret the observations, and in \S\ref{sec:discuss} we discuss some
further aspects of this discovery.

\section{Discovery \label{sec:disc}}

Unlike most L-dwarf candidates that are selected from infrared or deep
optical surveys, LSR0602+3910 (in further text LSR0602), was
identified as a new high proper motion star close to the Galactic
plane. The detection procedure relies on SUPERBLINK software
\citep{lepine1} -- an automated blink comparator working on DSS1 and
DSS2. After aligning the images from the two epochs, SUPERBLINK
subtracts them, thus revealing stars with significant motion. The
search in \citet{lepine1} concentrated on low Galactic latitudes and
stars with proper motions exceeding $0\farcs5 \, \yr^{-1}$, matching
the cut of the {\it Luyten Half-Second} (LHS) catalog \citep{lhs}. LHS
contains some 3600 stars with $\mu>0\farcs5 \, \yr^{-1}$, and is
fairly complete to $R_{\rm LHS}\sim 18.5$ in the northern skies and
away from the plane \citep{dawson,monet,flynn}. However, within
$10\degr$ of the plane, LHS stops being complete at $R_{\rm LHS}\sim
14$, because of the difficulties that Luyten had with identifying
moving stars in crowded fields. Here SUPERBLINK takes over. In
\citet{lepine1}, besides recovering 460 LHS stars, 141 new stars are
found with $13<R<20$. LHS is known to be a valuable source of nearby
stars, white dwarfs and subdwarfs. Adding 30\% of such stars should
add a similar percentage of the type of ``interesting'' stars present
in LHS. A spectroscopic campaign was undertaken to ascertain the
nature of these new stars. In \citet{lepine2} spectra were presented
for 105 stars, revealing many M dwarfs, and a number of subdwarfs and
white dwarfs. LSR1835+3259 was found to be only 6 pc away. The
spectrum of LSR0602 was not taken at that time. Another approach to
classifying proper motion stars and getting approximate photometric
distances employs the reduced proper motion diagram where the color is
an optical to infrared color (e.g. $V-J$). This method was described
in \citet{sgapjl} and was applied to the NLTT catalog \citep{nltt}. A
wide color baseline allows even crude photographic magnitudes to be
used. Matching the \citet{lepine1} sample to the 2MASS Second
Incremental Data Release (2IDR) produced 52 matches. Eleven stars
(most probably faint white dwarfs) did not have a match in 2IDR
areas. One object stood apart for its very red color. LSR0602 had
2MASS $J-K_s = 1.43$, indicating a possible L dwarf.

\section{Observations \label{sec:obs}}

\subsection{Spectroscopy}

The spectra of LSR0602 were obtained at the 3m Shane Telescope at Lick
Observatory, using the KAST spectrograph. We observed through a
$1\farcs5$ slit, and used the 600 lines/mm grating blazed at 7500\AA,
giving 6300-9100\AA\ coverage at 2.3\AA/pix. Dome flats were taken
after each telescope pointing to correct the fringing pattern of the
thinned CCD in the red. Comparison arc spectra were taken after each
pointing as well. Three 500 s exposures were obtained on 2002 Nov 28
with thin cirrus, and three 500 s exposures on Nov 30, of which one
was affected by clouds. Our effective exposure time is $\sim 2600$ s.
The 2D spectra are reduced and extracted using IRAF.  We have not
corrected for telluric absorption.  The individual spectra are
combined, weighted by the intensity. The combined spectrum, shown in
Figure \ref{fig:sp}, is flux-normalized, using the spectrum of the
spectroscopic standard BD +28 4211.

\subsection{Photometry}

During the same run, on 2002 Nov 27 and Dec 2, we obtained images of
LSR0602 on the 1m Nickel Telescope with Dewar\#2 CCD. The first night
was photometric, but the second was not. We derived zero points of the
photometric calibration for the first night, while the airmass
coefficients and linear and quadratic color-terms were derived from a
large number of Johnson-Kron-Cousins standards observed on Dec 3,
including the $V-I=4.000$ standard G 45-20. Calibration errors are
expected to be $\lesssim 0.02$ mag in all three bands ($VRI$). Since
the second night was not photometric, only relative photometry was
performed. Owing to the large number of stars in the field, the
extinction from clouds was well determined. Aperture corrections were
also determined precisely from a number of stars. Photometry is
summarized in Table \ref{tab:lsr}. Uncertainty in $V$ is dominated by
photon noise, in $R$ by the uncertain estimate of the cloud extinction
(since no $R$-band images of the field were obtained under photometric
conditions), and in $I$ from the uncertainty in $V-I$ color used to
convert from instrumental magnitudes.

\section{Analysis of LSR0602 \label{sec:ana}}

\subsection{Spectrum}

\begin{figure}
\plotone{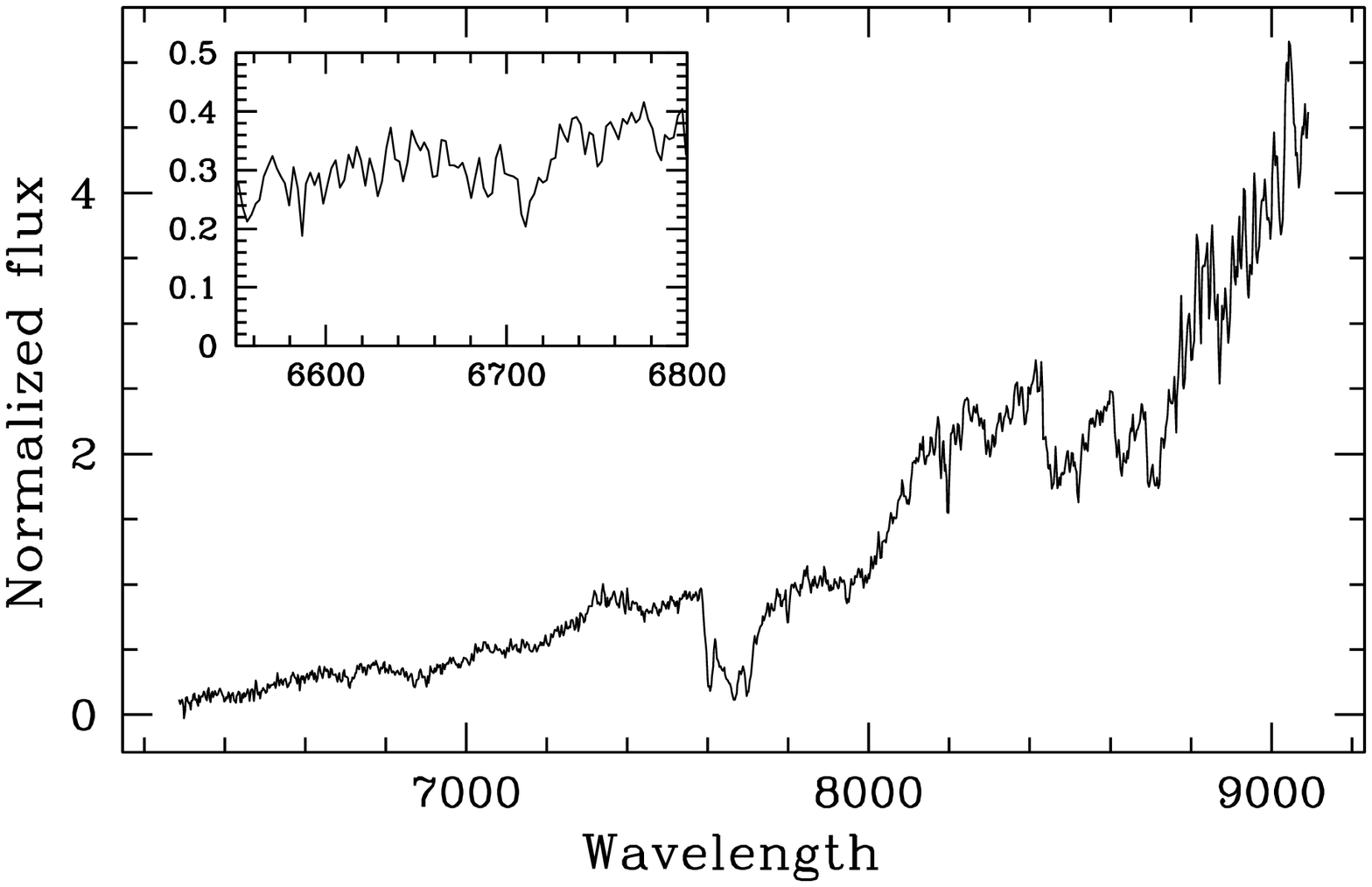}
\caption{Far-red spectrum of LSR0602, showing all features
characteristic of an early L dwarf. Total integration time was $\sim
2600$ s on the 3m Shane telescope.  Inset shows the region around the
lithium absorption line (Li I $\lambda6708$).\label{fig:sp}}
\end{figure}

We visually classify the spectrum of LSR0602 (Figure \ref{fig:sp}) as
type L1, based on templates and descriptions of L subtypes in
\citet{k99}. LSR0602 does not appear consistent with L0 because of the
strengths of Rb I and Cs I line doublets and the depth of the FeH
$\lambda8692$ band head. LSR0602 cannot be L2 because the TiO
$\lambda7053$ band head is still present, while the TiO $\lambda8432$
is of similar depth as the CrH $\lambda8611$ band head. The only
unusual feature for an L1 spectrum is the relatively narrow width of
the K I $\lambda\lambda7665/7699$ doublet.

The most intriguing feature of the LSR0602 spectrum is the detection
of lithium absorption (Li I $\lambda6708$). As can be seen from the
inset in Figure \ref{fig:sp}, the line seems to stand out, but the
region is noisy. The feature can also be seen in most individual
spectra. The centroid of the line in the combined spectrum lies at
6710\AA, 2\AA\ from the nominal position. Lines at longer wavelengths
mostly lie within 1\AA\ of their nominal positions. The dispersion
solution has an RMS value of 0.7\AA, so the ``offset'' of the Li I
line is most likely an artifact. The Li I line was recently
independently observed (K.\ Cruz \& I.\ N.\ Reid 2003, private
communication), confirming its brown dwarf status.

We measure the equivalent width (EW) of Li I to be $\sim7\,$\AA, which
is typical for L1 dwarfs \citep{k00}. According to \citet{k00}, only
10\% of L1 dwarfs have Li I, while 30\% exhibit ${\rm H}\alpha$
emission \citep{gizis00}. We do not detect ${\rm H}\alpha$. Since Li I
is visible only in younger L dwarfs ($\lesssim 1$ Gyr), while ${\rm
H}\alpha$ tends to be present in older ones \citep{gizis00}, the
absence of ${\rm H}\alpha$ in LSR0602 is consistent with it being a
brown dwarf. So are the low tangential and radial velocities,
indicating young disk membership.

\subsection{Brightness and Binarity}

According to Kirkpatrick's list, with a 2MASS magnitude of $K_s =
10.86$, LSR0602 comes second in brightness only to one other L dwarf:
2MASSW J0746425+200032 (or 2M0746). 2M0746 is one of the earliest L
dwarfs discovered \citep{reid00}, and has $K_s = 10.49$. Even in the
discovery paper, the trigonometric parallax of 2M0746 was available,
indicating an absolute magnitude $\sim 0.7$ mag brighter than the
``main sequence'', and thus a possible binary. This was confirmed by
\citet{reid01} who imaged 2M0746 with the {\it HST} Planetary Camera,
and found a companion a mere $0\farcs22$ away. The observations were
made in {\it HST} bands close to $V$ and $I$, so the model-dependent
extrapolation is needed to determine magnitudes of the components in
other bands. Using the estimated $\Delta K=0.36\pm0.20$ from
\citet{dahn}, the primary in 2M0746 acquires $K_s=11.08$, making it
$1.0\sigma$ fainter than LSR0602. The 2M0746 system does not show
lithium absorption (see Table \ref{tab:dist}), therefore it may, or
may not be substellar.

The question to ask is whether LSR0602 is also perhaps a close
binary. Our $I$-band image of LSR0602 taken in $1\farcs6$ seeing is
consistent with a point source. Since the magnitude difference between
the components would be smaller in $K$-band, we also examined 2MASS
images. While the FWHM of 2M0746 has a value which is $3\sigma$ larger
than that of the surrounding stars, LSR0602 has the same FWHM as other
stars. This does not rule out binarity, but puts stronger constraints
on it. Future observations with adaptive optics or optical
interferometers might find a companion, if one exists. The
determination of the trigonometric parallax would test LSR0602's
possible binarity as well.

While not surpassed in brightness among the brown dwarfs of L-type,
LSR0602 is fainter than the M9.5 brown dwarf LP944-20 \citep{tinney},
and comparable to the M8.5 brown dwarf 2MASSI J0335020+234235
\citep{reid02} and the L dwarf $\epsilon$ Ind B \citep{epsind}.

\subsection{Distance}

We determine the distance to LSR0602 in several ways.  Since we
measure the $I$ magnitude with a high precision, we can combine it
with the 2MASS $J$ to get $I-J=3.51$, and use the \citet{dahn}
$M_J(I-J)$ relation (calibrated with the largest available sample of
trigonometric parallaxes) to obtain a photometric distance. Thus we
get $d_{I-J}=10.4\pm1.1$ pc, where the error is due to the dispersion
in the $M_J(I-J)$ relation of 0.23 mag, and not from
$\sigma(I-J)\lesssim 0.04$. Another relation given by \citet{dahn}
relates spectral subclass to $M_J$. With it we get $d_{\rm
ST}=10.8\pm1.1$ pc, where again the error is from the 0.25 mag scatter
in the relation. That the two independent methods yield the same
distance estimate is quite reassuring. We have also derived a distance
from $J-K$ by asking that it on average (for the entire 250 L dwarf
sample) produce the same distance as the spectroscopic method. In this
way we get $d_{J-K}=10.0$ pc. However, it is well known that $J-K$
color is an inferior indicator of the absolute magnitude. The average
of the $I-J$ and spectroscopic distance estimates gives $d=10.6\pm0.8$
pc.

How does LSR0602 rank in distance compared to the other 250 L dwarfs?
To answer this, we compute the spectroscopic distances for the entire
sample.  The top of the list is shown in Table \ref{tab:dist}. We also
list trigonometric parallaxes if available. LSR0602 ranks eighth
judging from spectroscopic distances, however, due to binarity, 2M0746
actually lies farther.  In other cases, trigonometric distances are in
a very good agreement with spectroscopic estimates. Another five L
dwarfs have spectroscopic distances within $1\sigma$ of the LSR0602
estimate, but none of them has a trigonometric parallax. Of L dwarfs
estimated to be closer than LSR0602, all except one have optical
spectra, with none showing lithium.

\section{Sky distribution of L dwarfs and Conclusions \label{sec:discuss}}

\begin{figure}
\plotone{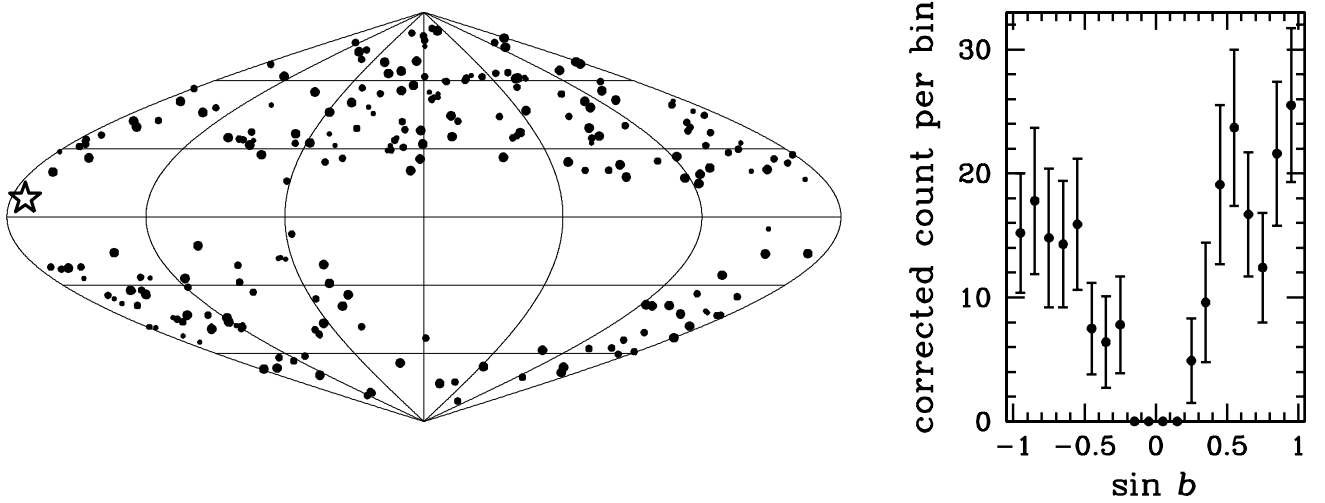}
\caption{Left panel: Map in Galactic coordinates (centered at the
Galactic center) showing 250 L dwarfs from the D.\ Kirkpatrick's
database. Different dot sizes correspond to different magnitude ranges
(see text). LSR0602 (star symbol at the left side) lies in the middle
of the ``zone of avoidance''. The absence of L dwarfs below the plane
reflects the 2MASS Second Incremental Data Release coverage. Right
panel: Distribution over Galactic latitude of $K_s<14.5$ L dwarfs
lying in the 2MASS 2IDR areas. Numbers are corrected for the fraction
of the sky that the 2IDR covers in the given latitude range. The
distribution is plotted against the sine of latitude to make bins (0.1
wide) correspond to equal areas on the sky.
\label{fig:map_count}}
\end{figure}

LSR0602 is a very bright L dwarf, yet it has escaped detection so
far. The reasons for this become obvious if we plot the distribution
of the 250 L dwarfs in the Galactic coordinate system (left panel of
Figure \ref{fig:map_count}). The Galactic center lies in the center of
the map, and L dwarfs are plotted as dots of three different
sizes. The largest ($K_s<13$), correspond to the limits of Sample A of
\citet{gizis00} and the bright \citet{k00} sample. Medium dots
($K_s<14.5$), correspond to the limits of \citet{k99,k00} and the
2MASS completeness limit. Fainter objects are plotted with small
dots. We notice that there is a ``zone of avoidance'' $\sim15\degr$
wide on each side of the plane. It is right there ($l=174\degr$,
$b=8\degr$), in the left corner of the map, where LSR0602 is located
(star symbol). Most searches for L dwarfs purposely avoided the region
around the Galactic plane in order to reduce the number of candidate
objects (mostly reddened stars). Another region that is conspicuously
devoid of L dwarfs is located below the plane. This region actually
corresponds to part of the sky around the south celestial pole that
was imaged only later by 2MASS, and is not a part of the 2IDR, where
most searches concentrated.

We can try to estimate the number of L dwarfs missed because searches
have been avoiding the plane. To make the estimate as free of bias as
possible, we select only L dwarfs lying within the 2MASS 2IDR
area. This amounts to 68\% of all L dwarfs, regardless of the
magnitude range. Since this percentage is considerably higher than the
2IDR sky coverage of 50\%, this confirms that most discoveries come
from 2IDR, and justifies our restriction to it. We now include only
$K_s<14.5$ objects, again in order to mimic the properties of searches
based on 2MASS. This leaves 133 L dwarfs. In order to obtain the
distribution of L dwarfs as a function of latitude, we want to correct
the raw counts per bin by the fraction of the sky in that bin covered
by the 2MASS 2IDR. Depending on the latitude range, this fraction
varies from 38\% to 67\%. Finally, in order to have each bin cover the
same area, we count in bins linear in $\sin b$. Thus each of the 20
bins covers $\sim 2000\, {\rm deg}^{2}$. The distribution is shown in
the right panel of Figure \ref{fig:map_count}. The error bars come
from the counting statistics. First, we confirm that there are no L
dwarfs in the $|b|<12\degr$ region. Had LSR0602 been discovered
previously (and it does lie in the 2IDR), it would have raised the
value in the $\sin b =0.15$ bin from 0 to $2.6\pm2.6$. We also notice
that the increase in the number is gradual towards the higher
latitudes, and levels off at $|b|\sim 30\degr$. There also seems to be
a statistically significant asymmetry between the northern and the
southern latitudes, perhaps a consequence of a less intensive
spectroscopic follow-up in the southern hemisphere. From the last five
bins (northern high latitude), we find the average number of 20 L
dwarfs per bin. This corresponds to 400 $(\pm40$) L dwarfs to
$K_s<14.5$ over the entire sky. However, the 2MASS coverage-corrected
number of known L dwarfs is 230 $(\pm20)$. Therefore some 40\% of L
dwarfs are missed in the current type of searches, mostly because the
Galactic plane region is avoided.

We present a discovery of a very bright L-type brown dwarf. Its space
velocity is consistent with the young disk. Since it is also one of
the closest L dwarfs, it should be included in a parallax
program. LSR0602 is an important target for follow-up
investigations. Due to its brightness, the blue part of the spectrum
and photometric variability can be easily studied. High resolution
imaging will be needed to investigate possible binarity. Finally, due
to its position in the sky that has been so far neglected, the
discovery indicates that many interesting nearby objects still await
discovery, especially using searches based on proper motion.

\acknowledgments This research program is being supported by NSF grant
AST-0087313 at the American Museum of Natural History, as part of the
NStars Program.  This publication makes use of VizieR and SIMBAD
Catalogue Services in Strasbourg, and data products from the Two
Micron All Sky Survey, which is a joint project of the University of
Massachusetts and the IPAC/Caltech, funded by the NASA and the NSF.

\clearpage

\clearpage

\begin{deluxetable}{lrll}
\tabletypesize{\footnotesize}
\tablecolumns{4} 
\tablewidth{0pt} 
\tablecaption{Data for LSR0602+3910. \label{tab:lsr}} 
\tablehead{ Datum & Value & Units & Reference\tablenotemark{a}}
\startdata 
RA (J2000) & 06 02 30.46 & h m s & 2\\
DEC (J2000) & +39 10 59.2 & d m s & 2\\
Gal.\ longitude & 173.86 & deg & 2\\
Gal.\ latitude & +8.16 & deg & 2\\
Epoch & 1998 Oct 19 & & 2\\
$\mu$       & 0.522 & arcsec yr$^{-1}$ & 3\\
p.a.         & 163.8 & deg & 3\\
$V$ & 20.88$\pm$0.12 & mag & 1\\
$R$ & 18.03$\pm$0.11 & mag & 1\\
$I$ & 15.80$\pm$0.03 & mag & 1\\
$J$ & 12.29$\pm$0.03 & mag & 2\\
$H$ & 11.46$\pm$0.03 & mag & 2\\
$K_s$ & 10.86$\pm$0.02 & mag & 2\\
Spectral Type & L1 V & & 1\\
Distance & 10.6$\pm$0.8 & pc & 1\\
$V_{\rm tan}$ &  $26\pm2$ & $\kms$ &  1+3 \\
$V_{\rm rad}$ &  $-5\pm28$ & $\kms$ &  1 \\
\enddata
\tablenotetext{a}{1 = This paper, 2 = 2MASS, 3 = \citet{lepine1}}
\end{deluxetable} 

\begin{deluxetable}{llrlrrr}
\tabletypesize{\footnotesize}
\tablecolumns{7} 
\tablewidth{0pt} 
\tablecaption{The Nearest L dwarfs \label{tab:dist}} 
\tablehead{ Name & Discovery reference & $K_s$\tablenotemark{a}
 & Sp.\ class\tablenotemark{b} & $d_{\rm ST}$ (pc) & $d_{\rm trig}$ (pc) 
& Li I EW (\AA)}
\startdata

DENIS-P J0255-4700 &	 \citet{martin} &	11.53 &	L8 V &	5.5 &	n/a &	$<1$ \\
2MASSW J1507476-162738 & \citet{reid00} &	11.30 &	L5 V &	7.3 &	7.3 &	$<0.1$ \\
2MASSW J0036159+182110 & \citet{reid00} &	11.03 &	L3.5 V & 7.8 &	8.8 &	$<0.1$ \\
GJ 1001B = LHS 102B & 	EROS (1999) &	11.40 &	L5 V &	8.3 &	9.6 &	no \\
2MASSI J0835425-081923 & Cruz (in prep.) &	11.14 &	L5 V & 8.6 & n/a & no \\
2MASSI J0746425+200032 & \citet{reid00} &	10.49 &	L0.5 V & 9.0 &	12.2 &	$<0.2$ \\
2MASSW J0045214+163445 & Wilson et al.\ (in prep.) & 11.37	& [L3.5?V] &	10.4 &	n/a &	? \\
LSR0602+3910 	& 	This paper	 &	10.86 &	L1 V &	10.8 &	n/a &	$\sim7$ \\
\enddata
\tablecomments{Based on L dwarf online archive maintained by D.\ Kirkpatrick 
(as of 2002 Dec 19), and notes from N.\ Lodieu (2003, priv.\ comm.) and K.\ 
Cruz (2003, priv.\ comm.)}
\tablenotetext{a}{From 2MASS.}
\tablenotetext{b}{Classification from D.\ Kirkpatrick archive, except for 
2MASS 0835-08 K.\ Cruz (2003, priv.\ comm.) and LSR0602.}

\end{deluxetable}

\end{document}